# A Residual Multi-task Network for Joint Classification and Regression in Medical Imaging

Junji Lin[a,b,†], Yi Zhang[a,†], Yunyue Pan[a], Yuli Chen[c], Chengchang Pan[a,*], Honggang Qi[a,*]

[a] School of Computer Science and Technology, University of Chinese Academy of Sciences, Beijing, 101408, China

[b] College of Surveying and GeoInformatics, Tongji University, Shanghai, 200092, China

[c] School of Computer Science, Shaanxi Normal University, Xi'an, Shaanxi, 710062, China

[†] Co-first author

[*] corresponding author, chpan.infante@qq.com, hgqi@ucas.ac.cn

**Abstract.** Detection and classification of pulmonary nodules is a challenge in medical image analysis due to the variety of shapes and sizes of nodules and their high concealment. Despite the success of traditional deep learning methods in image classification, deep networks still struggle to perfectly capture subtle changes in lung nodule detection. Therefore, we propose a residual multi-task network (Res-MTNet) model, which combines multi-task learning and residual learning, and improves feature representation ability by sharing feature extraction layer and introducing residual connections. Multi-task learning enables the model to handle multiple tasks simultaneously, while the residual module solves the problem of disappearing gradients, ensuring stable training of deeper networks and facilitating information sharing between tasks. Res-MTNet enhances the robustness and accuracy of the model, providing a more reliable lung nodule analysis tool for clinical medicine and telemedicine.

## 1    Introduction

Lung cancer is one of the most serious public health problems worldwide, resulting in approximately 1.8 million deaths per year [1-3]. At present, pulmonary nodules, as a key indicator of early diagnosis of lung cancer, are widely used in clinical screening [4,5]. Pulmonary nodules refer to circular or irregular lesions with a diameter less than or equal to 3cm in the lung, which can be divided into benign and malignant types [6,7]. Chest X-ray imaging is considered to be one of the effective means to detect pulmonary nodules [8]. Although existing methods have made remarkable achievements in the classification, detection and segmentation of single tasks, many models are still limited to processing a single task and lack the ability to simultaneously perform multi-task prediction and comprehensive analysis of multiple state indicators of nodules. Therefore, establishing models that can perform multiple tasks simultaneously and comprehensively evaluate nodule characteristics can bring better performance than learning them independently [9], which is of great significance for improving accuracy and efficiency in the early diagnosis of lung cancer.

To solve this problem, we propose a multi-task joint learning model based on deep learning, which combines multi-model fusion feature extraction, label smoothing, and dropout to improve lung nodule detection and classification performance. Contributions from this study include:

1. Multi-model fusion feature extraction: Integrating feature learning of different models enhances the ability of multi-task representation and significantly improves the overall performance.

2. Classification robustness optimization: By introducing label smoothing and dropout, the network's generalization ability and tolerance to class uncertainty are enhanced by randomly dropping some neurons.

3. Multi-task residual learning: By sharing feature layer and residual learning, task collaboration is optimized, information sharing between tasks is improved, and the comprehensive performance of the model is enhanced.

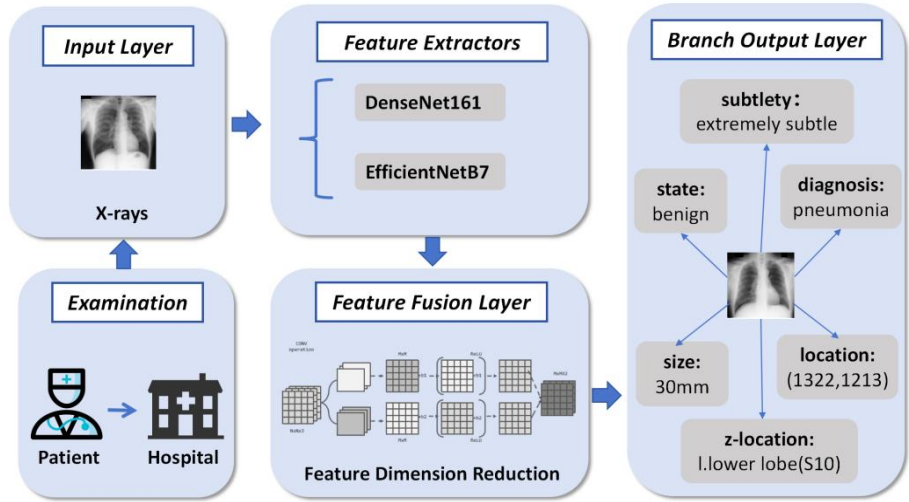

**Fig. 1.** This figure shows a multi-task residual learning model architecture for feature extraction and classification.

## 2      Related Work

In medical image analysis, deep learning techniques, particularly multi-task learning (MTL) and residual learning, have significantly enhanced diagnostic accuracy and efficiency. MTL has shown remarkable progress in various medical imaging tasks. For example, Amyar et al. improved pneumonia classification in COVID-19 CT detection by combining segmentation, classification, and reconstruction [10]. Duan et al. developed a double ventricular segmentation pipeline for cardiac MRI, enabling more accurate 3D modeling [11]. Samala et al. enhanced breast cancer diagnosis using multi-task transfer learning [12]. Chen et al. improved small object segmentation accuracy in glaucoma diagnosis by prepositioning objects in the feature layer [13]. Luo et al. enhanced myocardial segmentation and classification accuracy using a shared-weight independent encoder [14].

In automatic lung nodule recognition, MTL networks have also made significant strides. The YOLOv8-optimized model improved detection accuracy through enhanced feature fusion and loss function optimization [15]. The concept-res-v2 model addressed insufficient training data via transfer learning [16]. By integrating multi-model feature extraction and task-sharing mechanisms, these approaches provide more accurate and reliable tools for clinical pulmonary nodule detection.

## 3    Method

### 3.1    Feature extraction based on multi-model fusion

In this study, we propose a multi-task framework based on deep learning for the detection and classification of pulmonary nodules. The framework integrates feature extraction from multiple pre-trained models, such as DenseNet-161 and EfficientNet-B7, and adopts a shared multi-task network for multi-task learning. The overall structure of the framework is shown in Figure2.

First, the lung X-ray images are subjected to standard preprocessing, including resizing to 224x224 pixels, converting to tensors, and normalization. Then, the DenseNet-161 and EfficientNet-B7 models were used to extract the depth features to obtain the universal feature representation of the image. The features of the two models are extracted and connected respectively to form a comprehensive depth feature vector.

MultiTaskNet consists of a shared feature extraction layer and multiple task branches. The shared layer maps the input to a low-dimensional space and is processed by ReLU activation and Dropout. Tasks include: subtle classification, state classification, Z classification, diagnostic classification, X and Y regression, size regression, smoothing loss using labels, BCEWithLogits loss, and MSE loss. During the training process, the Adam optimizer (learning rate 0.0001) is used, the learning is guided by the weighted loss function, and the weights are updated by backpropagation. In the inference phase, the same pre-processing and multi-tasking is performed on each test image to generate the predicted results for each task. The framework improves diagnostic efficiency and accuracy through multi-task learning, and reduces overfitting problems.

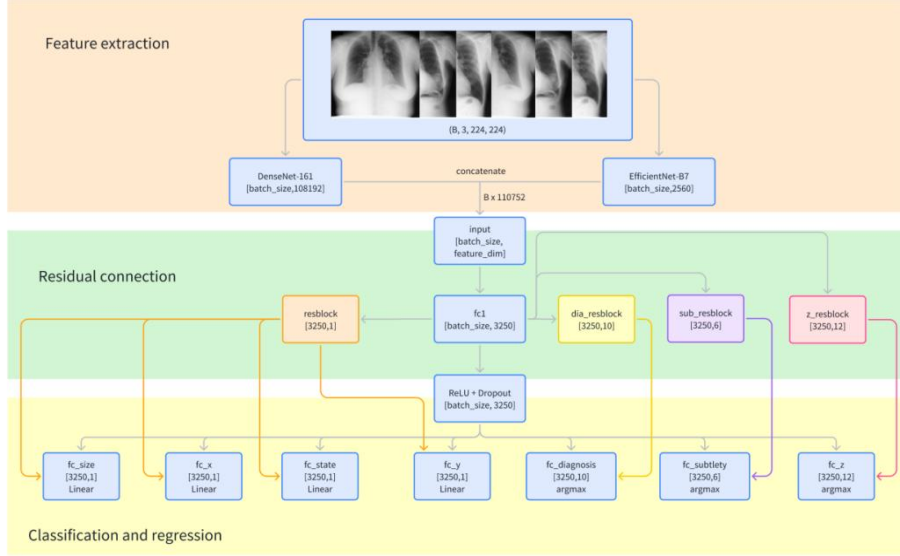

**Fig. 2.** This figure illustrates the detailed architecture of the multi-task residual learning model.

In medical image analysis, feature extraction is crucial for model performance. Traditional single pre-trained models struggle to capture both details and global information in complex tasks, limiting their application. To address this, a multi-model fusion strategy was proposed in this study, integrating features from different models to enhance their strengths, improve model expressiveness and robustness, and boost diagnostic performance while reducing overfitting [17].

Two pre-trained models, DenseNet-161 and EfficientNet-B7, were selected for feature extraction and fusion. DenseNet-161 enhances feature reusability through dense connections, improving detail capture and sensitivity for small nodule changes [18, 19]. EfficientNet-B7 optimizes the balance of depth, width, and resolution through composite scaling, enhancing performance with lower computational cost [20]. By combining DenseNet-161's detail capture with EfficientNet-B7's global feature learning, the model comprehensively captures various lung nodule characteristics, improving multi-task performance.

### 3.2    Multi-task learning

Our model design follows a multi-task learning framework to extract the general features of the image by sharing a fully connected layer, sharing the features of the specific output layer and then passing them to each task of the task for prediction, which include: Nodule concealability (subtle), nodule location (X, Y), nodule Size (Size), nodule State (State), nodule type (Z), diagnosis, which can solve multiple related tasks at the same time.

The model processes the image data through a shared feature extraction layer (such as fc1), and assigns the processed features to the output layer for each specific task. Such a multi-task learning structure enables the network to exploit the correlation between different tasks and improve the overall prediction accuracy.

### 3.3  Label smoothing

In traditional classification, cross-entropy loss with hard labels can lead to model overconfidence and reduced generalization, especially with noisy labels. Label smoothing addresses this by converting hard labels to soft labels, introducing uncertainty and improving tolerance to label noise [21, 22].

In this study, label smoothing was applied to multi-task learning, particularly for "subtle," "Nodule status," and "diagnosis" categories of pulmonary nodules. It effectively prevented overfitting, enhanced the model's ability to recognize subtle differences, and improved robustness and classification performance.

### 3.4  Residual connections

In this study, we introduce a residual module to enhance the multi-task learning framework. Residual learning uses skip connections to directly pass input information to subsequent layers, preventing vanishing gradients and accelerating network training [23]. We apply this module in the shared layer of the multi-task network, enabling each task to perform representation learning and iterative improvement based on shared features. This not only improves model performance for each task but also enhances knowledge sharing between tasks, allowing the network to leverage both global and local features. The residual module is defined as:

$$y=F(x)+x \qquad (1)$$

x is the input feature，F(x)is the feature representation after prior change，and y is the output of the residual module.With residual linkage, we can make predictions for each task based on the feature information extracted from all previous tasks, without having to learn the same features repeatedly.

## 4  Experiments and result

### 4.1  Datasets

The dataset used in this study is the JSRT (Japanese Society of Radiological Technology) chest X-ray lung nodule dataset, available on Kaggle at [https://www.kaggle.com/datasets/raddar/nodules-in-chest-xrays-jsrt]. It contains high-quality chest X-ray images with single lung nodules. The nodules are rated by 20 radiologists, with AUC values ranging from 0.72 to 0.89. Key features for model training include nodule subtlety, location, size, malignancy, state, and type.

### 4.2    lossFunction

In this experiment, the loss function is designed with different strategies to adapt to the classification and regression tasks in multi-task learning. Specifically, for the classification task, we use the Label Smoothing Loss, while for the regression task, we use the mean squared error Loss (MSE Loss).

The label smoothing loss is calculated as follows.

$$L_{LS} = -\sum_{c=1}^{C}[(1-\alpha)\cdot y_c \log(p_c) + \alpha \cdot \frac{1}{C}] \tag{6}$$

$y_c$ is the label of the target class，$p_c$ is the predicted probability of the model，C is the number of classes, and $\alpha$ is the smoothing factor.

For the state, x, y, and size tasks, we use the mean squared error loss (MSE):

$$L_{MSE} = \frac{1}{N}\sum_{i=1}^{N}(y_i - p_i)^2 \tag{7}$$

$y_i$ is the true value, $p_i$ is the value predicted by the model

The final total loss function is the weighted sum of the losses of all tasks, as follows.

$$L_{total} = \lambda_1 L_{subtlety} + \lambda_2 L_{state} + \lambda_3 L_z + \lambda_4 L_{diagnosis} + \lambda_5 L_x + \lambda_6 L_y + \lambda_7 L_{size} \tag{8}$$

Where, $\lambda_1, \lambda_2, \ldots, \lambda_7$ is the weight coefficient of each task

### 4.3    System Environment

In this study, all experiments were conducted under the following hardware and software environments: The operating system used was Ubuntu 18.04. The hardware configuration included two NVIDIA GeForce RTX 4090 GPUs (each with 24GB of video memory) and Intel(R) Xeon(R) Platinum 8352V CPUs operating at 2.10GHz, with a total of 144 CPU cores (36 cores per NUMA node). The software environment consisted of Python 3.9 and PyTorch 2.4.1.

### 4.4    Results  Classification and Regression task performance

In classification tasks for Subtlety, State, and Diagnosis, the model achieved high accuracy and F1 scores. For Stealthiness classification, accuracy was 0.8897 and F1 score was 0.8900, with good performance overall but room for improvement in recognizing extremely and very occult nodules. Presence state classification had an accuracy of 0.9485 and F1 score of 0.9482, indicating high accuracy in detecting nodules. Diagnostic classification showed accuracy of 0.8676 and F1 score of 0.8626, with some confusion between disease categories.

In regression tasks for nodule location and size prediction, performance was evaluated by MSE and MAE. For location prediction, MSE values for x and y coordinates were 46694.8984 and 13247.4424, and MAE values were 97.2524 and

A Residual Multi-task Network for Joint Classification and Regression in Medical Imaging   7

53.3991, respectively, with acceptable prediction errors. For size prediction, MSE was 14.0563 and MAE was 1.4045, indicating relatively accurate results.

### 4.5 Comparative experimental analysis

In this section, we will explore the results of comparative experiments in order to evaluate the effect of different nodule sizes on the performance of the model and the potential effect of different backbone networks on the performance of the lung nodule detection model. The experimental results in the supplementary table show that when our network Res-MTNet is used as the backbone network, the model shows the best performance on several evaluation indicators(the nodule size is projected to [0, 1] when calculating).

Table 1. Comparison of Classification Performance Across Different Models

| model | Subtlety | | State | | Diagnosis | | Location(Z) | |
|---|---|---|---|---|---|---|---|---|
| | Acc | F1 | Acc | F1 | Acc | F1 | Acc | F1 |
| resnet18 | 0.311 | 0.251 | 0.578 | 0.456 | 0.372 | 0.231 | 0.393 | 0.292 |
| resnet50 | 0.275 | 0.218 | 0.687 | 0.562 | 0.401 | 0.242 | 0.202 | 0.162 |
| Alexnet | 0.487 | 0.456 | 0.802 | 0.762 | 0.305 | 0.245 | 0.521 | 0.471 |
| Mobilenet | 0.805 | 0.798 | 0.965 | 0.964 | 0.444 | 0.419 | 0.563 | 0.517 |
| Googlenet | 0.451 | 0.428 | 0.652 | 0.582 | 0.361 | 0.218 | 0.464 | 0.374 |
| EfficientNetB7 | 0.382 | 0.352 | 0.617 | 0.479 | 0.338 | 0.264 | 0.448 | 0.429 |
| Res-MTNet | 0.875 | 0.875 | 0.911 | 0.912 | 0.860 | 0.856 | 0.882 | 0.884 |

Table 2. Comparison of Regression Performance Across Different Models

| model | x | | y | | Nodule Size | |
|---|---|---|---|---|---|---|
| | MSE | MAE | MSE | MAE | MSE | MAE |
| resnet18 | 0.03291 | 0.14568 | 0.02443 | 0.12653 | 0.000009 | 0.00255 |
| resnet50 | 0.03366 | 0.14573 | 0.03000 | 0.13492 | 0.000012 | 0.00251 |
| Alexnet | 0.01120 | 0.05064 | 0.00750 | 0.04084 | 0.000002 | 0.00096 |
| Mobilenet | 0.01316 | 0.06063 | 0.00738 | 0.04167 | 0.000002 | 0.00084 |
| Googlenet | 0.02320 | 0.11054 | 0.01701 | 0.10147 | 0.000011 | 0.00242 |
| EfficientNet-B7 | 0.03212 | 0.12658 | 0.01853 | 0.10152 | 0.000014 | 0.00270 |
| Res-MTNet | 0.01143 | 0.04673 | 0.00285 | 0.02417 | 0.000003 | 0.00085 |

### 4.6 Influence of different nodule sizes on model performance

Performance comparison experiments of different nodule sizes: The experimental data show the influence of different nodule sizes on the performance of the model. Here is the performance comparison by nodule diameter:

Table 3. Impact of Nodule Size on Classification Performance

| Size range(mm) | subtlety | | State | | Location(Z) | | Diagnosis | |
|---|---|---|---|---|---|---|---|---|
| | Acc | F1 | Acc | F1 | Acc | F1 | Acc | F1 |
| <10    | 1.000 | 1.000 | 0.900 | 0.897 | 0.900 | 0.867 | 1.000 | 1.000 |
| [10-20]| 0.861 | 0.861 | 0.907 | 0.906 | 0.883 | 0.882 | 0.872 | 0.873 |
| [20-30]| 0.866 | 0.854 | 0.900 | 0.908 | 0.766 | 0.770 | 0.900 | 0.912 |
| ≥30    | 0.900 | 0.895 | 1.000 | 1.000 | 0.900 | 0.893 | 0.800 | 0.801 |

**Table 4.** Impact of Nodule Size on Regression Performance

| Size range(mm) | x | | y | | Size | |
|---|---|---|---|---|---|---|
|   | MSE | MAE | MSE | MAE | MSE | MAE |
| <10    | 0.023034 | 0.064232 | 0.000898 | 0.182828 | 0.000001 | 0.000952 |
| [10-20]| 0.010723 | 0.444024 | 0.003113 | 0.024812 | 0.000002 | 0.000756 |
| [20-30]| 0.009272 | 0.048821 | 0.003308 | 0.028821 | 0.000001 | 0.000605 |
| ≥30    | 0.007110 | 0.040227 | 0.000319 | 0.009537 | 0.000229 | 0.002240 |

## 5     Conclusion

This paper proposes a deep learning framework based on multi-model fusion and multi-task learning for pulmonary nodule detection and classification. Features were extracted by DenseNet-161 and EfficientNet-B7, and combined with the label smoothing loss function to improve the generalization ability and robustness of the model. The experimental results show that the model performs well in multiple classification tasks, especially in the nodular presence classification task, achieving high accuracy and F1 scores.

However, the model still has some shortcomings: First, the feature processing is relatively simple, although two powerful pre-trained models are used, the diversity of feature fusion is still insufficient, and further optimization is needed. Secondly, the traditional classification model performs poorly in nodule location prediction, especially in precise positioning, which is far inferior to the target detection network. To improve accuracy, target detection algorithms such as Faster R-CNN or YOLO can be combined in the future. In addition, in the case of limited data sets, the generalization ability of the model may be limited, and the performance can be further improved by augmentation techniques or large-scale data sets in the future.

**Acknowledgments.** We thank all affiliates of School of School of Computer Science and Technology, University of Chinese Academy of Sciences for their valuable feedback.This work was financially supported by Natural Science Foundation of China (grant number 62271466).

**Disclosure of Interests.** The authors declare that there are no competing interests to report in the paper.

A Residual Multi-task Network for Joint Classification and Regression in Medical Imaging     9